\documentclass[journal]{IEEEtran}

\usepackage{amsmath}
\usepackage{amsfonts}
\usepackage{graphicx}
 \usepackage{multirow}
\usepackage{booktabs}
\usepackage{url} 

\title{Hydration Monitoring Using Urinary Biomarkers: A Hybrid Classical--Quantum Predictive Modeling Framework}

\author{
Sa\'ul Gonz\'alez-Bermejo$^1$,
Tommaso Albrigi$^1$, 
Borja V\'azquez-Morado$^1$,
Urko Regueiro-Ramos$^1$,\\
Daniel Casado-Faul\'\i$^1$,
Sergi C\`onsul-Pacareu$^1$,
Laia Alentorn$^2$,
Jordi Ferré$^2$,
Valentino Asole$^2$ \\
and Parfait Atchad\'e-Adelomou$^1$\\
(1) Lighthouse Disruptive Innovation Group S.L.\\
(2) Jungle Venture S.L.}

\begin{document}

\maketitle

\begin{abstract}

Hydration status is a key physiological indicator associated with cellular homeostasis, renal function, and overall health. Recent advances in smart sensing environments enable passive monitoring of urinary biomarkers that can provide continuous insight into hydration dynamics. In this work, we investigate predictive modeling approaches for hydration monitoring using biomarker data collected through the Predict Health Toilet (PHT) system. The problem is formulated as a regression task using urinary indicators such as urine specific gravity, conductivity, and volume. We evaluate classical machine learning models and quantum machine learning architectures based on variational quantum circuits. In particular, we introduce a modular Quantum Sequential Model (QSM) designed to construct flexible hybrid quantum–classical predictive pipelines. Experimental results compare classical regression models, symmetry-constrained quantum regressors, and QSM architectures. The results provide insights into the potential role of quantum machine learning in digital health monitoring systems and highlight the opportunities and current limitations of near-term quantum computing for physiological data analysis.

\end{abstract}


\section{Introduction}

Hydration is a fundamental physiological process regulating cellular homeostasis, thermoregulation, cardiovascular stability, and renal function. Even mild levels of dehydration can negatively affect cognitive performance, physical endurance, and metabolic balance. Chronic or recurrent dehydration has been associated with increased risks of hospitalization, kidney dysfunction, and cognitive decline, particularly among elderly populations \cite{popkin2010water,cheuvront2014hydration}.

Assessing hydration status in everyday environments remains a significant challenge. Traditional monitoring methods typically rely on episodic laboratory measurements, such as plasma osmolality tests or urine strip analysis \cite{armstrong2007urine}. While these approaches provide reliable physiological indicators, they are inherently reactive and require active clinical intervention. As a consequence, early dehydration signals may remain undetected between clinical assessments.

Recent advances in embedded sensing technologies and smart environments are enabling new paradigms for passive and continuous health monitoring \cite{park2020smart}. Within this context, smart sanitation systems have emerged as a promising platform for non-invasive physiological monitoring. Smart toilets, in particular, offer the unique possibility of integrating biomarker sensing into routine daily activities without requiring active user participation \cite{park2020smart}. These systems can capture urinary biomarkers such as urine specific gravity, conductivity, and volume, which provide valuable indicators of hydration status and renal concentration processes.

The \emph{Predict Health Toilet (PHT)} system follows this paradigm by transforming conventional sanitation infrastructure into a passive health monitoring platform capable of collecting physiological measurements during normal daily use. In this work, the system leverages urinary biomarker data obtained from the Smart Urinal platform, an IP-protected urinalysis solution that integrates spectrophotometry and electrochemical sensing to enable accurate, multi-parameter biochemical analysis directly at the point of use. This biosensing approach supports real-time, non-invasive, and non-supervised monitoring, allowing for continuous and scalable data collection in real-world settings. However, the effective interpretation of these biomarker signals requires predictive models capable of identifying relevant physiological patterns and translating them into clinically meaningful indicators.

Machine learning techniques have been widely applied to biomedical prediction tasks and have demonstrated strong performance when analyzing structured physiological datasets \cite{esteva2019guide}. These models are capable of capturing nonlinear relationships between biomarkers and physiological conditions, enabling early detection and risk prediction in digital health systems.

More recently, quantum machine learning (QML) has emerged as a novel research direction exploring the potential advantages of quantum computational models for pattern recognition and regression tasks \cite{schuld2015introduction,biamonte2017quantum}. Variational quantum circuits and hybrid quantum--classical learning architectures provide flexible frameworks for modeling nonlinear relationships within high-dimensional feature spaces while remaining compatible with noisy intermediate-scale quantum (NISQ) hardware.

Despite increasing interest in QML, its applicability to biomedical prediction problems remains largely unexplored. In particular, there is limited empirical evidence evaluating how quantum models compare with classical machine learning approaches when applied to physiological datasets derived from passive monitoring systems.

In this work, we investigate predictive modeling approaches for hydration monitoring using urinary biomarker data collected through the \emph{Predict Health Toilet (PHT)} system, a non-intrusive health monitoring platform designed to automatically capture physiological indicators from urine during routine bathroom use. By continuously collecting longitudinal biomarker measurements related to hydration and metabolic status, the system generates structured datasets suitable for predictive health analytics. Using these data, we evaluate both classical machine learning models and quantum machine learning architectures, including symmetry-constrained quantum regressors and a modular \emph{Quantum Sequential Model (QSM)} designed to enable the flexible construction of hybrid quantum--classical predictive pipelines.

The objective of this study is to analyze the potential role of quantum machine learning in digital health monitoring systems and to explore how emerging quantum architectures may complement classical predictive models in the analysis of physiological biomarker data.

\section{State of the Art}

Hydration assessment traditionally relies on physiological biomarkers such as plasma osmolality, urine osmolality, and urine specific gravity (USG) \cite{cheuvront2014hydration,armstrong2007urine}.

Urine specific gravity is widely used because it provides a practical proxy for urine concentration and hydration status \cite{armstrong2007urine}. Although USG correlates with urine osmolality, the two measures represent different physical quantities.

Recent work has explored the integration of biosensing technologies into smart sanitation infrastructure for health monitoring \cite{park2020smart}. These systems aim to transform everyday environments into platforms capable of collecting physiological data passively.

At the same time, machine learning methods have been widely applied in healthcare prediction and clinical decision support \cite{esteva2019guide}. Emerging research has also explored the use of quantum machine learning models for biomedical prediction tasks \cite{biamonte2017quantum}.

\section{Methodology}

This section introduces the physiological modeling framework used for hydration monitoring as well as the predictive learning formulation used to estimate hydration-related variables from urinary biomarkers.

\subsection{Biomarker Representation}

Hydration monitoring relies on physiological biomarkers that reflect the body's fluid balance and renal concentration processes. Urinary biomarkers have been widely used in both clinical medicine and sports physiology as practical indicators of hydration status \cite{cheuvront2014hydration,armstrong2007urine}.

Let the urinary biomarker feature vector observed at time $t$ be defined as

\begin{equation}
X_t =
\begin{bmatrix}
USG_t \\
C_t \\
V_t %\\
%pH_t
\end{bmatrix}
\label{eq:biomarker_vector}
\end{equation}

where

\begin{itemize}

\item $USG_t$ denotes urine specific gravity
\item $C_t$ denotes urine conductivity
\item $V_t$ denotes urine volume
%\item $pH_t$ denotes urine acidity

\end{itemize}

The vector defined in (\ref{eq:biomarker_vector}) captures complementary biochemical properties related to hydration status. Urine specific gravity (USG) provides a proxy measurement of urine concentration, as it provides a rapid estimate of urine, while conductivity reflects ionic content within the urine \cite{armstrong2007urine}. Urine volume provide contextual physiological signals that contribute additional information about renal function and fluid balance.

These biomarkers have been widely used in hydration assessment studies due to their ability to reflect short-term changes in body fluid regulation \cite{cheuvront2014hydration,popkin2010water}. For example, clinical literature typically considers values between $1.000$ and $1.040$ as the physiological operating range for USG-based hydration assessment.

\subsection{Predictive Learning Problem}

The objective of the predictive modeling framework is to estimate hydration-related physiological variables from urinary biomarker measurements.

Let the dataset be defined as

\begin{equation}
D = \{(X_i,y_i)\}_{i=1}^{N}
\label{eq:dataset}
\end{equation}

where $X_i$ denotes the biomarker vector defined in (\ref{eq:biomarker_vector}) and $y_i$ represents the target variable to be predicted.

The predictive task consists of learning a function

\begin{equation}
f:\mathbb{R}^d \rightarrow \mathbb{R}
\label{eq:model_function}
\end{equation}

such that

\begin{equation}
\hat{y}_i = f(X_i)
\label{eq:prediction}
\end{equation}

minimizes the empirical loss function

\begin{equation}
\mathcal{L} =
\frac{1}{N}
\sum_{i=1}^{N}
(y_i-\hat{y}_i)^2
\label{eq:mse}
\end{equation}

The loss function defined in (\ref{eq:mse}) corresponds to the mean squared error, which is widely used in regression tasks in biomedical prediction problems \cite{esteva2019guide}.

This formulation allows the model to learn nonlinear relationships between urinary biomarkers and hydration-related physiological variables.

In practice, several classical machine learning models can be used to approximate the function $f(\cdot)$, including linear regression models, kernel-based methods, and neural networks. More recently, quantum machine learning models based on variational quantum circuits have also been proposed as alternative regression architectures capable of learning complex nonlinear mappings in high-dimensional feature spaces \cite{schuld2015introduction,biamonte2017quantum}.

\section{Dataset and Feature Engineering}

This section describes the preprocessing pipeline applied to the biomarker dataset prior to model training. Feature engineering plays a critical role in both classical and quantum machine learning models, as it directly influences model stability, generalization performance, and computational feasibility.

The dataset used in this work consists of urinary biomarker measurements collected from passive monitoring events. Each observation corresponds to a biomarker vector defined in (\ref{eq:biomarker_vector}) together with associated contextual and physiological variables.

\subsection{Feature Standardization}

Physiological variables are measured in different physical units and may exhibit significantly different numerical ranges. Without normalization, variables with larger numerical magnitudes can dominate the optimization process and bias the learning algorithm.

To mitigate this issue, all features are standardized using z-score normalization

\begin{equation}
x'_j =
\frac{x_j - \mu_j}{\sigma_j}
\label{eq:standardization}
\end{equation}

where $\mu_j$ and $\sigma_j$ denote the empirical mean and standard deviation of feature $j$, respectively.

Standardization ensures that each feature contributes proportionally during model training and improves numerical stability during optimization. This preprocessing step is widely used in machine learning pipelines for biomedical data analysis \cite{bishop2006pattern,hastie2009elements}.

\subsection{Dimensionality Reduction}

In many biomedical datasets, correlations exist between physiological variables. For example, urine specific gravity and urine conductivity both reflect urine concentration and therefore may exhibit strong statistical dependencies.

Dimensionality reduction techniques are commonly applied in order to reduce redundancy between correlated variables while preserving the underlying structure of the data \cite{jolliffe2016pca}.

In this work, Principal Component Analysis (PCA) is used to project the feature space onto a lower-dimensional subspace

\begin{equation}
\tilde{x} = P^T x
\label{eq:pca_projection}
\end{equation}

where $P$ represents the matrix of principal components derived from the covariance matrix of the dataset.

The covariance matrix of the standardized dataset is defined as

\begin{equation}
\Sigma =
\frac{1}{N}
\sum_{i=1}^{N}
(X_i - \mu)(X_i - \mu)^T
\label{eq:covariance_matrix}
\end{equation}

where $\mu$ denotes the empirical mean vector.

The eigenvectors of $\Sigma$ define the principal component directions, while the associated eigenvalues $\lambda_j$ represent the variance captured by each component.

The number of retained components $k$ is selected such that

\begin{equation}
\sum_{j=1}^{k}\lambda_j
\ge
\alpha
\sum_{j=1}^{d}\lambda_j
\label{eq:pca_variance}
\end{equation}

where $\alpha$ represents the desired variance preservation ratio.

In this work, $\alpha$ is typically chosen between $0.95$ and $0.99$, ensuring that most of the variance of the original dataset is retained while reducing the dimensionality of the feature space.

\subsection{Dimensionality Constraints for Quantum Models}

Dimensionality reduction becomes particularly important when training quantum machine learning models.

Current noisy intermediate-scale quantum (NISQ) hardware is limited by the number of available qubits and by noise levels present in quantum devices \cite{preskill2018nisq}. Consequently, the number of features that can be encoded into a quantum circuit is typically constrained by the number of available qubits.

By applying PCA prior to quantum feature encoding, the dimensionality of the feature space can be reduced to a number compatible with current quantum hardware while preserving most of the information contained in the original dataset.

This preprocessing step therefore plays a crucial role in enabling hybrid classical–quantum learning pipelines.

\section{Dataset and Data Sources}

The experimental evaluation of the proposed models relies on multiple complementary data sources combining physiological biomarker measurements, population-level hospital statistics, and structured physiological feature definitions. The integration of these heterogeneous datasets enables the construction of predictive models capable of linking individual hydration biomarkers with population-level health outcomes.

\subsection{Predict Health Toilet (PHT) Biomarker Dataset}

The primary physiological dataset originates from the Predict Health Toilet (PHT) sensing platform, which continuously captures urinary metrics over time through an embedded sensing system integrated into a smart sanitation environment. This gives an array of measurements, per each time window $t$, for each sensor system used.

The PHT system records a set of urinary measurements associated with hydration status, including:

\begin{itemize}
\item Conductivity sensor signals
\item Optical spectral sensor (8-band)
%\item Flow rate sensor
\item Temperature and Environmental Signals
%\item Urine pH
\end{itemize}

These measurements are collected during natural user interactions with the system and therefore enable passive and non-invasive health monitoring.

%\subsection{Hospital Morbidity Dataset}

%To contextualize the physiological measurements within broader health outcomes, this study also considers population-level hospitalization data derived from the Spanish \textit{Encuesta de Morbilidad Hospitalaria} (Hospital Morbidity Survey) maintained by the Instituto Nacional de Estadística (INE).

%The survey provides statistical information on hospital admissions classified according to the International Classification of Diseases (ICD-10-MC). The dataset captures the frequency of hospitalizations associated with specific diagnoses and demographic groups.

%The data used in this study span the period

%\begin{equation}
%2010 \leq t \leq 2024
%\end{equation}

%and cover the geographic region of Catalonia.

%The dataset includes the following variables:

%\begin{itemize}
%\item ICD-10 diagnostic code
%\item Diagnostic description
%\item Year of admission
%\item Month and week of admission (when available)
%\item Gender
%\item Age group
%\item Health region
%\item Hospital sector
%\item Hospital identifier (CNH code)
%\item Total number of diagnoses per category
%\end{itemize}

%These variables enable the construction of epidemiological indicators related to dehydration-associated hospitalizations.

%The use of aggregated hospital statistics allows the modeling framework to explore potential correlations between physiological hydration signals and population-level hospitalization trends.

\subsection{Physiological Feature Dictionary}

Additional physiological context is provided by the Phase 2 Feature Dictionary, which defines structured descriptors of urinary biomarkers and their physiological interpretations.

The dictionary organizes physiological measurements into standardized feature definitions, including:

\begin{itemize}
\item Biomarker measurement types (e.g. ideal water reference and ground truth measurements)
\item Physiological interpretation ranges (e.g. age, gender)
\item Derived hydration indicators (e.g. absorbance/reflectance, conductivity variability as a function of electrode immersion depth)
\item Data normalization parameters (e.g. perceptual color space representations)
\end{itemize}

This structured representation facilitates consistent feature engineering across heterogeneous datasets.

In particular, the dictionary enables the alignment of physiological variables collected through the PHT platform with clinically interpretable hydration indicators, represented as a biomarker feature vector in Eq~\ref{eq:biomarker_vector}.

\subsection{Data Integration Strategy}

The datasets used in this study operate at different observational scales:

\begin{itemize}
\item \textbf{Physiological scale:} individual urinary biomarker measurements
\item \textbf{Clinical scale:} physiological feature definitions and biomarker interpretations
%\item \textbf{Population scale:} hospitalization statistics associated with dehydration-related diagnoses
\end{itemize}

The modeling pipeline integrates these data sources through a feature engineering process that maps physiological observations to predictive urine biomarkers while preserving the epidemiological context of dehydration-related health outcomes.

This multi-source data integration framework supports the development of predictive models that combine physiological sensing with population-level health indicators.

\section{Computational Framework}

\subsection{Hybrid Classical–Quantum Infrastructure}

The experimental evaluation of the predictive models was conducted using a hybrid computational infrastructure combining classical cloud-based machine learning environments with a quantum experimentation platform.

The classical machine learning models were implemented and trained using Microsoft Azure Machine Learning (Azure ML). Azure ML provides a scalable cloud environment for data preprocessing, model training, and evaluation, enabling the deployment of standard machine learning pipelines and regression models. In this work, Azure ML was used to develop and evaluate the classical predictive models based on urinary biomarker features.

In contrast, the quantum machine learning experiments were conducted using the UniQuE experimentation platform \cite{casado2025unique}. UniQuE is a hybrid research framework designed to support rapid prototyping of quantum and hybrid quantum--classical machine learning models within controlled experimental pipelines.

Within this study, UniQuE was specifically used to:

\begin{itemize}
\item implement variational quantum circuit models,
\item evaluate structured quantum regressors,
\item implement the Quantum Sequential Model architecture,
\item perform comparative experiments between quantum and classical predictive models.
\end{itemize}

The separation of classical and quantum experimentation environments allows the evaluation of quantum models under controlled experimental conditions while maintaining a robust classical baseline implemented in a production-grade machine learning infrastructure.

This hybrid experimental setup ensures both scalability for classical model training and flexibility for exploring emerging quantum machine learning architectures.

\section{Quantum Machine Learning Framework}

The predictive modeling pipeline integrates classical machine learning methods with quantum machine learning (QML) models based on variational quantum circuits. The objective is to investigate whether quantum feature representations can capture relevant physiological patterns present in urinary biomarker data \cite{schuld2015introduction,biamonte2017quantum}.

Given the dimensionality constraints of noisy intermediate-scale quantum (NISQ) hardware \cite{preskill2018nisq}, QML models operate on the reduced feature vector obtained via PCA (cf. \eqref{eq:pca_projection}--\eqref{eq:pca_variance}). The QML framework adopted in this work consists of: (i) quantum feature encoding, (ii) variational learning, (iii) measurement, and (iv) hybrid classical optimization.

\subsection{Quantum Feature Encoding}

After dimensionality reduction, the resulting feature vector $x \in \mathbb{R}^k$ is encoded into a quantum state using angle embedding. The encoding unitary is defined as

\begin{equation}
U_{enc}(x) =
\prod_{j=1}^{k}
R_Y(x_j),
\label{eq:angle_embedding}
\end{equation}

where $R_Y(\theta)$ denotes a rotation gate around the $Y$ axis. The corresponding encoded state is

\begin{equation}
|\psi(x)\rangle =
U_{enc}(x)|0\rangle^{\otimes k}.
\label{eq:quantum_state}
\end{equation}

This encoding maps classical features into the Hilbert space of $k$ qubits, enabling nonlinear representations through subsequent entangling operations \cite{schuld2015introduction}.

\subsection{Variational Quantum Circuits}

Learning is performed using parameterized quantum circuits (variational quantum circuits, VQCs). A generic VQC model can be expressed as

\begin{equation}
U(x,\theta) =
\prod_{l=1}^{L}
V_l(\theta_l)\,U_{enc}(x),
\label{eq:vqc_model}
\end{equation}

where $V_l(\theta_l)$ denotes trainable circuit blocks and $\theta$ represents the set of trainable parameters.

Predictions are obtained via the expectation of an observable $O$,

\begin{equation}
\hat{y} =
\langle
\psi(x,\theta)
|
O
|
\psi(x,\theta)
\rangle,
\label{eq:quantum_prediction}
\end{equation}

with $|\psi(x,\theta)\rangle = U(x,\theta)|0\rangle^{\otimes k}$. Hybrid optimization is performed using classical optimizers such as Adam \cite{kingma2017adam}, typically implemented in differentiable quantum programming frameworks \cite{bergholm2022pennylane}.

\subsection{Data Re-uploading}

To increase expressivity under shallow-depth constraints, the model incorporates data re-uploading \cite{perezsalinas2020datareuploading,atchade2023fourier}. In this architecture, the encoding map is injected multiple times across layers:

\begin{equation}
U(x,\theta) =
\prod_{l=1}^{L}
U_{enc}(x)\,V_l(\theta_l).
\label{eq:data_reupload}
\end{equation}

Data re-uploading increases functional approximation capacity without requiring deep circuits, which is critical under NISQ noise regimes \cite{preskill2018nisq}.

\subsection{Quantum Sequential Model (QSM)}

While symmetry-constrained models (cf. Section~\ref{sec:structured_su_models}) introduce strong inductive biases, they may limit systematic architectural exploration. To enable a more flexible and modular design, we adopt the Quantum Sequential Model (QSM), a layer-wise compositional framework inspired by sequential constructions in classical deep learning \cite{bishop2006pattern,hastie2009elements}.

A QSM is defined as an ordered composition of quantum layers:

\begin{equation}
U_{\mathrm{QSM}}(x,\theta) =
\mathcal{L}_K(\theta^{(K)}) \circ \cdots \circ
\mathcal{L}_2(\theta^{(2)}) \circ
\mathcal{L}_1(x,\theta^{(1)}),
\label{eq:qsm_definition}
\end{equation}

where each $\mathcal{L}_k$ is a unitary transformation acting on the qubit register and $\theta=\{\theta^{(1)},\ldots,\theta^{(K)}\}$ denotes the complete trainable parameter set.

In practice, QSM layers can implement: (i) data encoding, (ii) trainable variational blocks, (iii) entangling patterns, and (iv) measurement schemes. This modularity allows substituting encoding maps, entanglers, and heads without redesigning the full model.

\subsubsection{Variational Blocks and Entanglement}

A generic trainable block used within QSM can be expressed as:

\begin{equation}
V(\theta)=
\left(\prod_{q=1}^{k}
R_X(\theta_{q,1})R_Y(\theta_{q,2})R_Z(\theta_{q,3})\right)
\,U_{\mathrm{ent}},
\label{eq:qsm_variational_block}
\end{equation}

where $U_{\mathrm{ent}}$ denotes an entangling operator (e.g., ring-CNOT, fully-connected CNOT, or other hardware-efficient patterns). This design corresponds to hardware-efficient ans\"atze commonly used in NISQ learning pipelines \cite{preskill2018nisq,schuld2015introduction}.

\subsubsection{Measurement and Quantum Latent Representation}

Let $\{O_j\}_{j=1}^{m}$ denote a set of observables. The measured latent quantities are:

\begin{equation}
z_j(x)=\langle O_j\rangle=
\langle 0|U^\dagger_{\mathrm{QSM}}(x,\theta)\,
O_j\,
U_{\mathrm{QSM}}(x,\theta)|0\rangle.
\label{eq:qsm_measurements}
\end{equation}

The final prediction is obtained from the quantum latent vector $z(x)=(z_1,\ldots,z_m)$ via a classical head:

\begin{equation}
\hat{y}=g\big(z_1(x),\ldots,z_m(x)\big),
\label{eq:qsm_head}
\end{equation}

where $g(\cdot)$ can be the identity (direct regression), a linear map, or a shallow neural regressor. This hybrid design allows flexible output calibration while keeping the quantum model focused on feature representation.

\subsection{Structured Quantum Regressors via Symmetry Constraints}
\label{sec:structured_su_models}

In addition to QSM, we explore structured regressors constrained by symmetry-inspired unitary families. Unitary operators acting on $n$ qubits lie in the group

\begin{equation}
U \in SU(2^n),
\label{eq:su_group}
\end{equation}

where $SU(2^n)$ is the special unitary group of degree $2^n$ \cite{hall2015liegroups}. Symmetry-constrained circuits can act as architectural regularizers by restricting the accessible dynamical subspace, potentially improving trainability and interpretability at the cost of reduced flexibility.

\section{Experimental Results}\label{sec:experimental_results}

This section reports the experimental protocol and comparative evaluation of (i) classical regression baselines, (ii) structured symmetry-based quantum regressors (SU-family), and (iii) the Quantum Sequential Model (QSM) introduced in \eqref{eq:qsm_definition}.

\subsection{Models Evaluated}

\emph{Classical baselines.} We evaluate standard regression models widely used in biomedical prediction pipelines (specifically using XGBRegressor \cite{Chen_2016,xgboostPython}, based on gradient boosting algorithms), trained using the objective in \eqref{eq:mse}. This training processes where fit using Hyperopt library \cite{bergstra2012makingsciencemodelsearch} to find the optimal hyperparameters.

\emph{QML--SU models.} We evaluate symmetry-constrained quantum regressors with low qubit count (e.g., SU(3) on 2 qubits) that impose structured inductive biases via restricted unitary families \cite{hall2015liegroups}.

\emph{Quantum Sequential Model (QSM).} We evaluate the modular QSM architecture defined in \eqref{eq:qsm_definition}, using angle embedding \eqref{eq:angle_embedding}, variational blocks \eqref{eq:qsm_variational_block}, and measurement-based heads \eqref{eq:qsm_head}. The QSM uses data re-uploading \eqref{eq:data_reupload} to increase expressivity \cite{perezsalinas2020datareuploading}.

\subsection{Training Protocol}

All models are trained on the same training split and evaluated on the same held-out test split. For QML models, optimization is performed in a hybrid loop (quantum forward pass + classical update) using Adam \cite{kingma2017adam} implemented through a differentiable quantum framework \cite{bergholm2022pennylane}. Dimensionality reduction via PCA (cf. \eqref{eq:pca_projection}--\eqref{eq:pca_variance}) is applied before quantum encoding to satisfy NISQ constraints \cite{preskill2018nisq}.

\subsection{Evaluation Metrics}

We report three complementary evaluation axes, following a homogeneous methodology across models:

\emph{Explained variance.} The coefficient of determination:
\begin{equation}
R^2 =
1 -
\frac{\sum_{i=1}^{N}(y_i-\hat{y}_i)^2}
{\sum_{i=1}^{N}(y_i-\bar{y})^2}.
\label{eq:r2}
\end{equation}

\emph{Average accuracy.} Mean absolute error:
\begin{equation}
MAE =
\frac{1}{N}\sum_{i=1}^{N}|y_i-\hat{y}_i|.
\label{eq:mae}
\end{equation}

\emph{Tolerance accuracy.} The proportion of predictions within clinically meaningful tolerance margins:
\begin{equation}
\mathrm{Acc}_{\delta} =
\frac{1}{N}\sum_{i=1}^{N}\mathbb{I}\left(|y_i-\hat{y}_i|\le \delta\right),
\quad \delta \in \{1,2,3\},
\label{eq:tolerance_accuracy}
\end{equation}
where $\mathbb{I}(\cdot)$ is the indicator function.

In addition, we analyze error distributions via residuals $\varepsilon_i=y_i-\hat{y}_i$ to assess robustness and outlier sensitivity.

\subsection{Architectural Comparison: Resources and Inductive Properties}

To synthesize methodological and architectural differences between quantum families, we report a two-level comparison: (i) quantum resources and (ii) inductive/expressive properties.

\begin{table}[t]
\caption{Architectural and resource comparison between QML families.}
\label{tab:qml_resources}
\centering
\begin{tabular}{@{}lll@{}}
\toprule
\textbf{Characteristic} & \textbf{SU(3) -- 2 Qubits} & \textbf{QSM} \\
\midrule
Number of qubits & 2 & Variable (typically 8) \\
Embedding type & Symmetry-structured & Angle embedding \\
Data re-uploading & Yes & Yes \\
Variational blocks & Symmetry-restricted & Hardware-efficient ansatz \\
Entanglement pattern & Symmetric Ising interactions & CNOT (ring / generic) \\
Trainable parameters & Low--moderate & Moderate--high \\
Circuit depth & Limited by symmetry design & Scales with \#layers \\
Scalability in qubits & Limited & High (layer-controlled) \\
Noise sensitivity & Moderate & Depth-dependent \\
NISQ compatibility & High & High \\
\bottomrule
\end{tabular}
\end{table}

\begin{table}[t]
\caption{Inductive bias and expressivity comparison between QML families.}
\label{tab:qml_inductive}
\centering
\begin{tabular}{@{}lll@{}}
\toprule
\textbf{Property} & \textbf{SU(3) -- 2 Qubits} & \textbf{QSM} \\
\midrule
Physical inductive bias & High & Low \\
Symmetry restriction & Explicit & Unrestricted \\
Interpretability & High & Moderate \\
Architectural flexibility & Low & Very high \\
Architecture search capacity & Limited & Broad \\
Structural modifiability & Reduced & High \\
Functional expressivity & High under symmetry & High unrestricted \\
Over-parameterization risk & Low & Moderate \\
Ease of training & High & Depth-dependent \\
Transferability across tasks & Limited & High \\
\bottomrule
\end{tabular}
\end{table}

\subsection{Quantitative Comparison}

We report results for each prediction task (e.g., USG, conductivity, volume) using the metrics in \eqref{eq:r2}--\eqref{eq:tolerance_accuracy}. Classical baselines generally provide strong performance, while quantum models show competitive behavior under reduced feature spaces. In particular, SU-models may benefit from strong inductive biases under limited resources, whereas QSM provides a flexible hypothesis space that can scale with circuit layers.

\begin{table*}[t]
\caption{Predictive performance comparison across classical and quantum models for hydration biomarker prediction.}
\label{tab:prediction_results}
\centering
\begin{tabular}{lcccccc}
\toprule
\textbf{Task} & \textbf{Model} & \textbf{$R^2$} & \textbf{MAE} & \textbf{Accuracy $\pm1$} & \textbf{Accuracy $\pm2$} & \textbf{Accuracy $\pm3$} \\
\midrule

\multirow{3}{*}{USG Prediction}
 & Classical Regression & 0.91 & 1.69 & 0.41 & 0.75 & 0.80 \\
 & QML SU Model & 0.52 & 4.57 & 0.18 & 0.29 & 0.44 \\
 & QSM Model & 0.84 & 2.35 & 0.29 & 0.54 & 0.75 \\

\midrule

\multirow{3}{*}{Urine Conductivity}
 & Classical Regression & 0.88 & 1.74 & 0.49 & 0.69 & 0.81 \\
 & QML SU Model & 0.49 & 4.78 & 0.16 & 0.28 & 0.42 \\
 & QSM Model & 0.67 & 3.61 & 0.14 & 0.37 & 0.48 \\

\midrule

\multirow{3}{*}{Urine Volume$^{*}$}
 & Classical Regression & 0.90 & 22.41 & 0.66 & 0.94 & 0.96 \\
 & QML SU Model & - & - & - & - & - \\
 & QSM Model & 0.89 & 26.54 & 0.61 & 0.89 & 0.95 \\

\bottomrule
\end{tabular}
\end{table*}

\begin{table*}[t]
\caption{Summary comparison between classical models, symmetry-based quantum models, and the Quantum Sequential Model (QSM).}
\label{tab:model_comparison}
\centering
\begin{tabular}{lccc}
\toprule
\textbf{Property} & \textbf{Classical ML} & \textbf{QML SU} & \textbf{QSM} \\
\midrule

Model flexibility & High & Medium & Very high \\
Inductive bias & Low & High & Moderate \\
Parameter count & Moderate & Low & Moderate--high \\
Scalability & High & Limited & High \\
Interpretability & Moderate & High & Moderate \\
Hardware constraints & None & Low qubit requirement & Depends on depth \\
Noise sensitivity & Low & Moderate & Depth dependent \\

\bottomrule
\end{tabular}
\end{table*}

We recommend reporting a consolidated table with rows = tasks and columns = (Classical, SU-QML, QSM-QML) for $R^2$, $MAE$, and $\mathrm{Acc}_{\delta}$ at $\delta=\{1,2,3\}$.

$^{*}$Volume accuracy metrics are computed using $\delta=\{25,50,75\}$, due to the comparatively wide range of target data, to provide a similar assessment of model performance with Conductivity and USG predictions.

\subsection{Error Distribution Analysis}

Residual distributions $\varepsilon_i=y_i-\hat{y}_i$ are analyzed to characterize robustness and outlier sensitivity. This analysis complements aggregate metrics by revealing whether errors are concentrated near zero or dominated by high-variance tails, which is common in physiological measurements.

\subsection{Visual Comparison}

\begin{figure}
    \centering
    \includegraphics[width=0.8\linewidth]{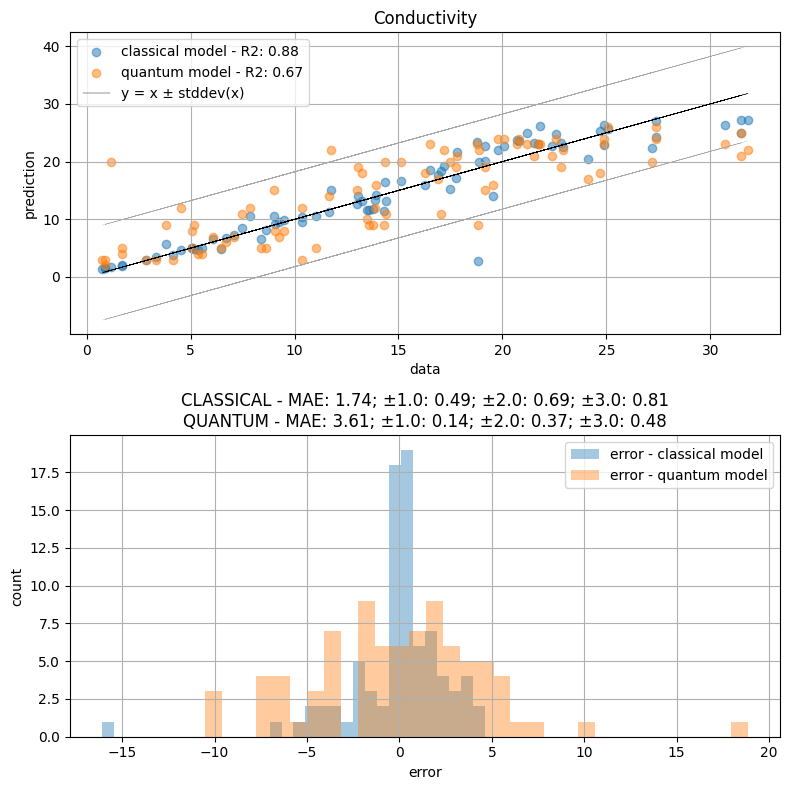}
    \caption{Visual model performance comparison on the Conductivity dataset. Top: true data on the x-axis, classical (blue) and quantum (orange) predictions on the y-axis; bottom: error distribution (same palette)}
    \label{fig:conductivity}
\end{figure}

\begin{figure}
    \centering
    \includegraphics[width=0.8\linewidth]{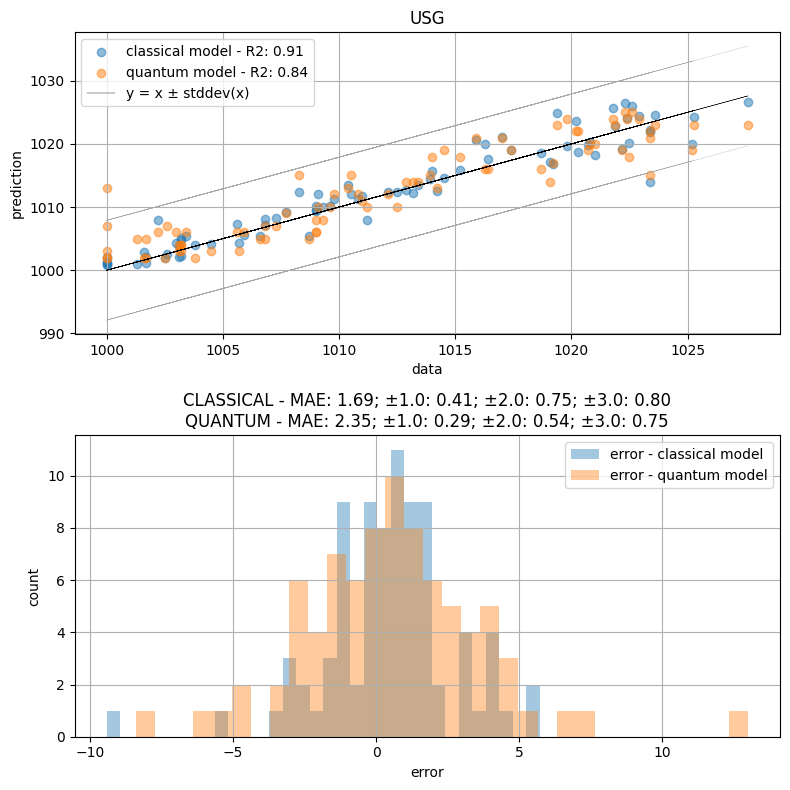}
    \caption{Visual model performance comparison on the USG dataset. Top: true data on the x-axis, classical (blue) and quantum (orange) predictions on the y-axis; bottom: error distribution (same palette)}
    \label{fig:usg}
\end{figure}

\begin{figure}
    \centering
    \includegraphics[width=0.8\linewidth]{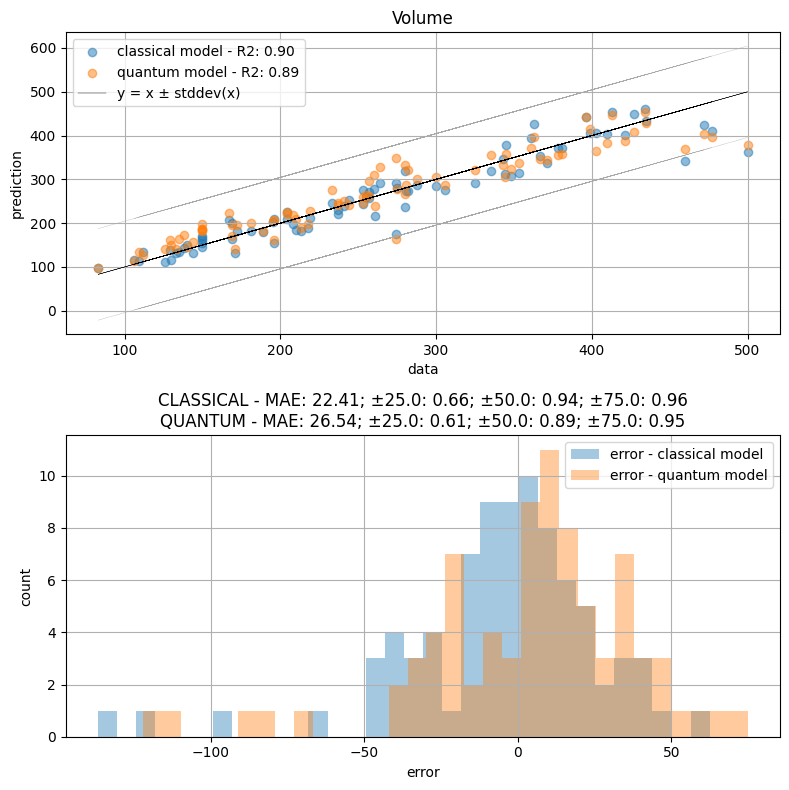}
    \caption{Visual model performance comparison on the Volume dataset. Top: true data on the x-axis, classical (blue) and quantum (orange) predictions on the y-axis; bottom: error distribution (same palette)}
    \label{fig:volume}
\end{figure}

A visual comparison of the performance of the classical and quantum models that achieve the best performance and their error distribution is provided in Figures~\ref{fig:conductivity}~\ref{fig:usg}~\ref{fig:volume}.

\section{Discussion}\label{sec:discussion}

The experimental results presented in Section~\ref{sec:experimental_results} provide several insights into the applicability of quantum machine learning models for physiological prediction tasks based on urinary biomarkers. The quantitative comparisons reported in Tables~\ref{tab:prediction_results} and \ref{tab:model_comparison} summarize the predictive performance obtained across the evaluated modeling approaches.

First, the results confirm that classical machine learning models currently provide strong predictive performance for the hydration-related variables analyzed in this study. As shown in Table~\ref{tab:prediction_results}, regression models trained on urinary biomarker features achieve robust predictive accuracy. This observation is consistent with the existing literature in biomedical data analysis, where classical regression and ensemble methods often perform well on structured physiological datasets.

The evaluation of quantum models reveals several interesting characteristics. Structured quantum regressors based on symmetry-constrained unitary transformations exhibit stable learning behavior even when implemented with a small number of qubits. The results summarized in Table~\ref{tab:prediction_results} indicate that these models maintain consistent predictive performance despite the limited representational capacity of shallow quantum circuits. This stability can be attributed to the inductive biases introduced by symmetry constraints, which effectively reduce the hypothesis space explored during training. Nevertheless, their performance is still low compared to the classical and QSM models. Volume performance metrics are not reported as the model failed to consistently predict target data.

While these models provide robustness and interpretability, their architectural rigidity limits the exploration of more expressive functional representations. In contrast, the Quantum Sequential Model (QSM) introduced in this work provides a more flexible architectural framework that enables the systematic construction of hybrid quantum–classical models through sequential composition of encoding layers, variational blocks, and measurement heads.

The modular design of the QSM architecture allows exploration of larger hypothesis spaces while maintaining compatibility with noisy intermediate-scale quantum (NISQ) devices. As illustrated in Table~\ref{tab:prediction_results}, the QSM architecture provides a more flexible modeling strategy capable of capturing nonlinear relationships between physiological biomarkers and hydration states through combinations of angle embedding, data re-uploading, and hardware-efficient variational layers.

An important aspect of the present study is that all quantum model evaluations were conducted using quantum circuit simulators. While simulation provides a controlled environment for benchmarking model architectures, it does not fully capture the computational characteristics of real quantum hardware. As quantum processors continue to improve in terms of qubit counts, coherence times, and gate fidelities, executing these models on actual quantum devices may enable the exploitation of intrinsic quantum properties such as entanglement and quantum interference.

Consequently, the current results should be interpreted as an initial exploration of quantum modeling approaches rather than a definitive assessment of their performance limits. The relatively modest performance differences observed between classical and quantum models (Tables~\ref{tab:prediction_results} and \ref{tab:model_comparison}) may partly reflect the constraints of simulation-based experimentation and the limited circuit depths compatible with current NISQ architectures.

From a biomedical perspective, the integration of predictive models within the Predict Health Toilet (PHT) system opens interesting opportunities for passive health monitoring. By combining biomarker acquisition, predictive modeling, and digital health interfaces, the system enables continuous assessment of hydration status without requiring active user intervention.

Although the present study focuses on hydration-related biomarkers, the proposed modeling framework can be extended to additional physiological indicators, potentially enabling early detection of metabolic, renal, or systemic health conditions.

Future work will focus on three main directions: the evaluation of the proposed quantum architectures on real quantum hardware, the exploration of deeper and more expressive quantum circuits as hardware capabilities improve, and the validation of the predictive framework on larger physiological datasets collected through the PHT sensing infrastructure.

\section{Conclusion}

This work presented a predictive modeling framework for passive hydration monitoring based on urinary biomarkers collected through the Predict Health Toilet (PHT) sensing infrastructure. The proposed approach integrates physiological biomarker acquisition with machine learning and exploratory quantum machine learning models within a hybrid computational pipeline.

Experimental results demonstrate the feasibility of using urinary sensor measurements for predictive hydration assessment and support the potential of smart sanitation environments as platforms for non-invasive preventive health monitoring.

Future research will focus on large-scale clinical validation, improved physiological feature modeling, and further investigation of hybrid quantum--classical approaches for digital health prediction systems.

\section{Code Availability}

The code used to reproduce the experiments presented in this study is publicly available on GitHub:

\begin{center}
\url{https://github.com/Lighthouse-DIG/PredictHealthToilet-PHT}
\end{center}

Classical models were implemented using Microsoft Azure Machine Learning, while the quantum machine learning models were implemented within the UniQuE experimentation framework and evaluated using quantum circuit simulators.

\section*{Acknowledgment}

The Subvencions RETECH program supported this research. %The authors thank the Instituto Nacional de Estadística (INE) team for providing access to the datasets used in this study.

\newpage
\bibliographystyle{unsrt}
\bibliography{references}

\end{document}